\documentclass[
reprint,
amsmath,amssymb,
aps, superscriptaddress,
prb
]{revtex4-2}

\usepackage{graphicx}% Include figure files
\usepackage{dcolumn}% Align table columns on decimal point
\usepackage{bm}% bold math
\usepackage{amsmath}
\usepackage{mathrsfs}
\usepackage{mathtools}
\usepackage{xcolor}
\usepackage{xspace}
\usepackage{orcidlink}

\newcommand{\sg}{SIM-GRAPH\xspace}
\newcommand{\ket}[1]{|#1\rangle}

\newcommand{\fiteq}[2][0.88]{\resizebox{#1\columnwidth}{!}{$#2$}}
\usepackage{bbold}

\begin{document}

\preprint{APS/123-QED}

\title{SIM-GRAPH: A universal guide to symmetric interactions}
%SIM-GRAPH 2.0: A complete guide

\author{R.~C. Verstraten\orcidlink{0000-0002-4386-5210}}%
% \affiliation{$^1$Institute for Theoretical Physics, Utrecht University, Princetonplein 5, 3584CC Utrecht, The Netherlands}
\author{C. Morais Smith\orcidlink{0000-0002-4190-3893}}%
\affiliation{Institute for Theoretical Physics, Utrecht University, Princetonplein 5, 3584CC Utrecht, The Netherlands}

\date{\today}

\begin{abstract}
Symmetry plays a central role throughout physics, from Fourier analysis on discrete lattices to the classification of elementary particles in the standard model. In finite quantum systems, symmetries are often discrete, such as reflection and rotational symmetries. Here, we present the SIM-GRAPH method (Symmetric Ising Models -- Graph Reduction And Projected Hamiltonians)~\cite{verstraten2025control}, which uses such symmetries to efficiently calculate ground-state observables of interacting quantum systems. By projecting all interactions onto a symmetric subset of the system, the method effectively reduces the number of sites, thus providing an exponential speedup over standard exact diagonalization. Furthermore, we introduce two extensions that broaden the applicability of the method and allow for a controlled trade-off between computational cost and accuracy.
%One of the foundation of physics lies in symmetry: The inability to distinguish between differing situations. Its utilization has provided us with tools like Fourier analysis on discrete lattices up to the very foundations of the standard model. On finite size systems or even molecules, we often find discrete symmetries such as mirror and rotational translations. Here, we provide a complete introduction how to utilize these symmetries to efficiently calculate ground-state observables of interacting quantum systems based on the SIM-GRAPH method: Symmetric Ising Models -- Graph Reduction And Projected Hamiltonians~\cite{verstraten2025control}. The core idea is to project all interactions to a symmetric subset, effectively reducing the number of sites. Depending on the amount of symmetry, this reduction tends to provide a significant exponential speedup compared to exact diagonalization. Further more, we present two extensions of the method, increasing its range of applicability and introducing the flexibility to vary between speed and accuracy.
\end{abstract}

\maketitle

%\tableofcontents

\section{\label{sec:introduction}Introduction}

Symmetries may be broadly classified as continuous or discrete, as well as global or local, and play a central role in the description of quantum many-body systems. Symmetry principles underlie the emergence and classification of phases of matter, including phenomena such as spontaneous symmetry breaking and collective behavior in interacting systems~\cite{anderson1972more,wen2004quantum}. In quantum mechanics, symmetry is associated with transformations that leave the Hamiltonian invariant and thereby constrain the structure and dynamics of physical systems~\cite{weyl1952symmetry,wigner2012group,tung1985group,tinkham2003group}. In this framework, symmetries are represented by (anti)-unitary operators acting on the Hilbert space, leading to selection rules, degeneracies, and conserved quantities. A fundamental result is provided by Noether's theorem, which yields a correspondence between continuous symmetries and conservation laws, such as energy and momentum conservation arising from time- and space-translation invariance, respectively~\cite{noether1971invariant,bajardi2022noether}. While continuous symmetries give rise to conserved quantities, discrete symmetries instead impose constraints on the spectrum and eigenstates, as exemplified by lattice translational symmetry leading to a crystalline momentum defined modulo the reciprocal lattice vectors.

Magnetic materials provide a versatile platform for investigating interacting quantum many-body systems, where localized spins coupled via exchange interactions give rise to a wide range of emergent phenomena, including magnetic order, frustration, and exotic quantum phases~\cite{sachdev2008quantum,nolting2009quantum,coey2010magnetism,balents2010spin,anderson1987resonating}. A key feature of such systems is quantum entanglement, which captures non-classical correlations between subsystems and provides a unifying framework for characterizing complex many-body states~\cite{amico2008entanglement,horodecki2009quantum,eisert2010colloquium,laflorencie2016quantum}. In particular, entanglement has proven instrumental in understanding quantum criticality and the scaling behavior of correlations near phase transitions~\cite{osterloh2002scaling,osborne2002entanglement,vidal2003entanglement,calabrese2004entanglement,dagotto1994correlated}. The interplay between many-body quantum interactions and entanglement thus governs both the equilibrium and dynamical properties of magnetic systems, and underlines their relevance for quantum simulation and quantum information processing~\cite{georgescu2014quantum,bravyi2017complexity}.

The complexity of interacting quantum many-body systems, which arises from the exponential growth of the Hilbert space with system size, constitutes a central obstacle in their theoretical description. For a system of $N$ local degrees of freedom with an on-site Hilbert space of dimension $d$, the fully interacting Hilbert space scales as dimension $d^N$, which rapidly renders exact treatments infeasible~\cite{feynman2018simulating}. A standard approach is to represent the Hamiltonian as a matrix acting on the full Hilbert space and compute its spectral properties using exact diagonalization (ED). Standard algorithms typically scale as $\mathcal{O}(L^3)$ for an $L\times L$ matrix~\cite{demmel2008performance}, but the exponential scaling $L\sim d^N$ implies an overall computational cost that grows exponentially with system size, both in runtime and memory requirements. In practice, this restricts ED to relatively small systems, even when exploiting sparsity and symmetries through block diagonalization into irreducible representations and symmetry-adapted bases~\cite{schmitz2020quantum,ghassemi2025simultaneous}. To overcome these limitations, a broad range of numerical and analytical techniques has been developed, targeting physically relevant states that occupy only a small, structured subset of the full Hilbert space. Prominent examples include stochastic methods such as quantum Monte Carlo (QMC), tensor network approaches such as matrix product states (MPS) and projected entangled pair states (PEPS), and renormalization-based techniques including the density matrix renormalization group (DMRG)~\cite{sandvik2010computational,thijssen2000computational,bauer2011alps,saad2011numerical,lanczos1950iteration,orus2014practical,white1992density,schollwock2011density,verstraete2008matrix,fehske2007computational}. These methods leverage the entanglement structure and locality of interactions to achieve efficient representations in specific regimes. However, each comes with intrinsic limitations, such as the treatment of real-time dynamics, limits in system size, or convergence issues with high degeneracies. As a result, the development of alternative frameworks capable of capturing the essential structure of complex quantum systems remains an active area of research.

Here, we further develop the hybrid framework \sg: \textit{Symmetric Ising Models --- Graph Reduction And Projected Hamiltonians} proposed earlier~\cite{verstraten2025control}. Although not exact, this framework can have significant computational benefits compared to existing methods. The first application of this method, presented in Ref.~\cite{verstraten2025control}, already demonstrated a significant speedup while computing the ground-state average magnetization of a transverse field Ising model (TFIM) on a 24-site Sierp\`inski triangle. In this model, ED required a month server time for 50 points, QMC required two weeks server time for 50 points, variational mean-field (VMF) required 3 hours for 10000 points on a strong desktop, whereas on a simple laptop \sg took under 30 seconds for 10000 points. There are three key principles to the effectiveness of the \sg framework. Firstly, we generate effective Hamiltonians whose spectrum mimics that of the original one, but have a significantly smaller Hilbert space. Secondly, we do not aim at computing the entire wavefunction. Instead, we only seek enough information to compute local observables. The final key insight is that the wavefunction of the effective Hamiltonian contains sufficient information to compute those observables due to the symmetry constrains of the original Hamiltonian. 

In this work, we present two new generalizations of \sg, both of which remain exponentially faster than ED. The first generalization improves the accuracy of \sg for many different geometries by retaining more of the most likely superpositions within the groundstate. Particularly the results describing even geometries and systems with strong long-range interactions greatly benefit from this extension. This accuracy can even be increased, at the tradeoff of using more computational resources. The second generalization is based on partial symmetries. For instance, here we might allow for states which break mirror symmetry but retain rotational symmetry. Although this is again at the slight cost of computational speed, we find that certain geometries, including frustrated ones, are extremely well described by this approximation. 

This article is organized as follows. In Sec.~\ref{sec2}, we introduce the working principles behind the \sg method. In Sec.~\ref{sec3}, we apply \sg to the TFIM and show the challenge of different type of geometries. In Sec.~\ref{sec4}, we introduce \sg 2, which is a generalization based on partial reductions. In Sec.~\ref{sec5}, we introduce \sg 3, which is a generalization based on partial symmetries. In Sec.~\ref{sec6}, we present some toy models and explicitly compare the differences between the results using \sg and exact methods. In Sec.~\ref{sec7}, we explore a large variety of geometries and compare the convergence of the different \sg methods with exact results. Finally, we present our conclusions in Sec.~\ref{sec:con}.

%%%%%%%%%%%%%%%%%%%%%%%%%%%%%%%%%%%%%%%%%%%%%
%%%%%%%%%%%%%%%%%%%%%%%%%%%%%%%%%%%%%%%%%%%%%

\section{Working principles of \sg}\label{sec2}
 One of the working foundations of physics is to use symmetry in a problem to find a good basis  to solve the equations. Translation symmetric problems are often solved in a Fourier basis and rotational symmetric problems are computed in polar or spherical coordinates. These type of transformations help to decouple the different degrees of freedom without altering the original system. These symmetries are described using group theory, with many different applications~\cite{tung1985group}. The problem with quantum interactions, however, is that they can spontaneously break symmetries~\cite{strocchi2020symmetry}. For instance, consider two interacting spin-1/2 particles and let us represent spin up by 1 and spin down by 0. They may find themselves in a symmetric state consisting of $\ket{\psi}=\ket{10}+\ket{01}$. If it was not for spontaneous symmetry breaking, we could apply these symmetry principles directly to a quantum wavefunction and reduce the degrees of freedom in a similar fashion. Therefore, the challenge to use symmetry in interacting quantum systems is that the symmetry is only partially endowed on the wavefunction. Nevertheless, we know that ground-state observables are not affected by the spontaneous symmetry breaking. Therefore, we will not focus on computing the wavefunction, but on computing observables. Then, we may alter the wavefunction, as long as it does not affect the observable of interest. To ensure that the outcome remains relevant to the original system, we will only allow changes which do not alter the eigenenergy of the state.

 As a general strategy, we will consider a mapping between interacting Hamiltonians $H$ and graphs $G$, and we will analyze the symmetry before constructing the effective Hamiltonian:
 \begin{equation}
      H \to G \to G_{SIM} \to H_{SIM}.
 \end{equation}
To analyze the symmetry within the Hamiltonian, we convert its terms into a \textit{graph} $G$, where each node $i$ represents a site, and each connection $C_{ij}$ represents a term of the Hamiltonian, weighted by its interaction strength. Note that there may be self-loops $C_{ii}$ in this graph, representing local terms. Then, we analyze the graph using group theory, and find the \textit{automorphism group} $S$. This group consists of the cycles between symmetric sites, and its generator(s) $\hat{S}_i$ define the symmetries of $H$, such as rotating or mirroring in some way. For more complex geometries, one may even use improved algorithms to compute the symmetry group~\cite{shah2025efficient}. Next, we treat the sites modulo the group cycles. For this, it is easiest to select \textit{representatives} $S_i$ from each cycle that lie in the same sector, although this is not required. In our code, we select the most bottom-left point from each cycle, matching our counting order of the site labels. The \textit{symmetry reduced graph} $G_{SIM}$ is produced by projecting all connections to the representative sites and removing the other sites,
\begin{equation}
    G_{SIM}= P_C[G],\; \text{ where }\;
    P_C: \: C_{ij} \mapsto C_{S_i S_j}.
\end{equation}
Remark that the labeling of the sites may be changed in this step, since some low-index sites may be removed in the projection, and we want to retain a counting order $1,2,\dots$ in the symmetric-site labels.

\begin{figure}[t]
    \centering
    \includegraphics[width=\columnwidth]{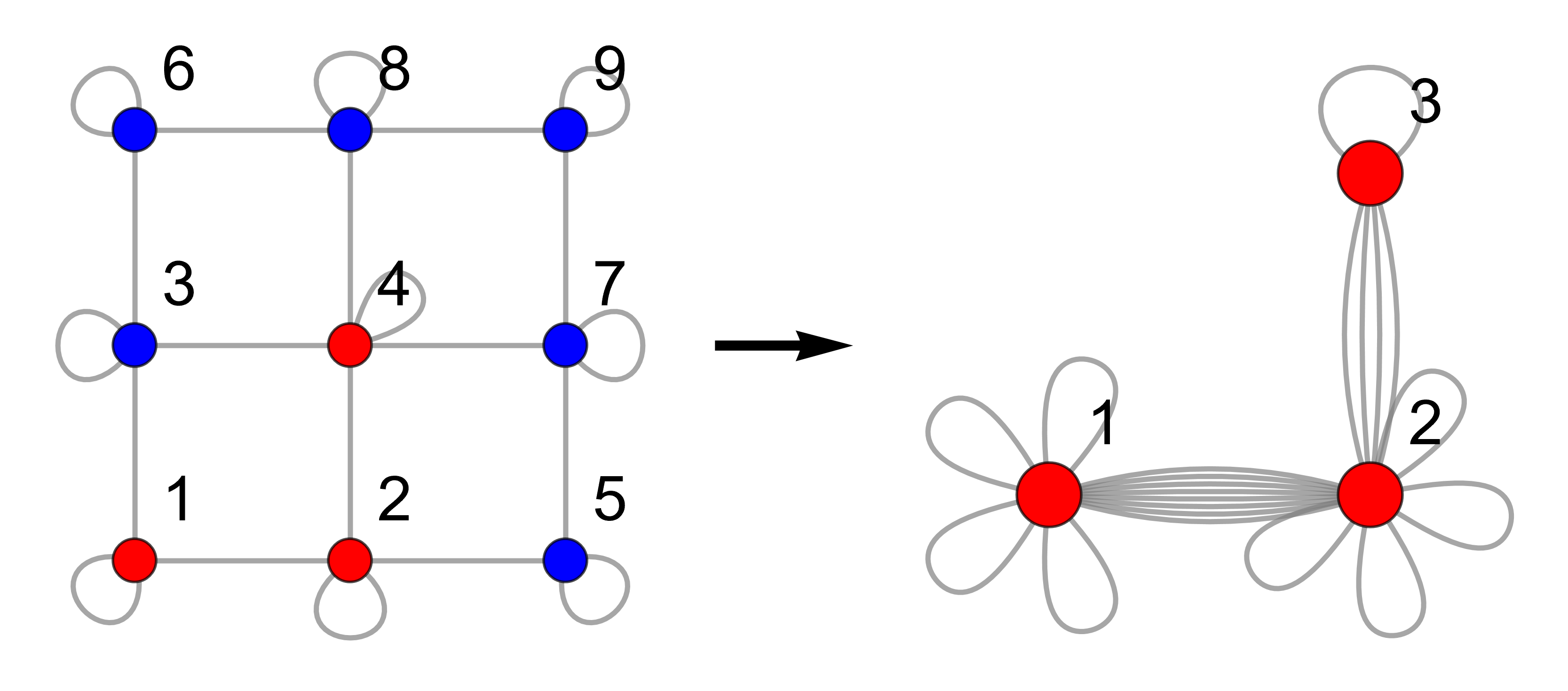}
    \caption{Graph reduction of a 3x3 square lattice with NN (links) and local (loops) interactions. The symmetry group is generated by a 90 degree rotation along site 4. The three distinct symmetric representations $S_i$ are highlighted in red on the left hand side. On the right hand side, we see the reduced graph, where all interactions are projected to the symmetric subset and the other sites are removed. The remaining sites are relabeled to a counting order.}
    \label{fig:squarereduction}
\end{figure}

As an example, a reduction of a 3x3 square lattice is presented in Fig.~\ref{fig:squarereduction}. Here, the symmetry group consists of the cycles $S=\{(1,6,9,5),(2,3,8,7),(4)\}$, where we find three symmetric representations on sites 1, 2, and 4. In more common terms, we see that the square can be split up into 4 corners, 4 edges, and a center point. Let us also consider the group actions. Clockwise rotation of the square is achieved by moving every cycle one to the right (e.g. $(1,6,9,5)\mapsto (5,1,6,9)$). Mirroring along the vertical axis is achieved by reversing the first cycle and switching the second and fourth term in the second cycle (e.g. $(2,3,8,7)\mapsto(2,7,8,3)$). Combinations of these two actions can generate all 8 possible configurations of the square. After the projection, we can indeed observe that the total number of lines remains unchanged by the graph reduction.

Gathering all terms in $G_{SIM}$, we can now construct the reduced Hamiltonian $H_{SIM}$. We find that we can use the graph projection in a very similar way to define a direct projection $P_S$ on the Hamiltonian, which depends on the symmetry group $S$. Here, each site index $i$ in the Hamiltonian is replaced by its symmetric representation $S_i$,
\begin{equation}
    H_{SIM}= P_{S}\;\hat{S}^{-1} H \hat{S}\; P_{S},\; \text{ where }\;
    P_S: i\mapsto S_i
\end{equation}
and $\hat{S}$ is the symmetry transformation from site $i$ to $S_i$.
Note that $P_S$ is indeed a projection, as $P_S^2[i]=P_S[S_i]=S_{S_i}=S_i=P_S[i]$. Effectively, we see that all terms in the Hamiltonian are retained, but they are transferred to the symmetric sector. This yields an effective Hamiltonian which has a much lower number of sites, roughly by a factor of the symmetry type (e.g. 2 for mirror symmetry, 3 for triangular rotational symmetry, and 2x3=6 for triangular plus mirror symmetry). The number of sites $N$ exponentially determines the dimension of the Hilbert space based on the number of local states. In the case of the spin-1/2 Ising model, that is $d=2^N$. In turn, $d$ is the most important factor to determine the complexity of finding the ground-state wavefunction. Thus, a linear reduction in $N_{SIM}\sim N/|S|$ yields an exponential reduction in $d$, which in turn speeds up computational time exponentially. In principle, any known numerical technique (e.g., QMC, Mean-field, or MPS) may be used to compute the effective properties of $H_{SIM}$, but in this paper we will only use exact diagonalization. 
Symmetry in observables combined with conservation of interactions lie at the heart of \sg. 
We also note that this symmetry reduction is different from methods which split the geometry into a sum of smaller geometries, known as cluster decomposition~\cite{lyons1964cluster,schmidt2021correlated}, which allow for a tiling of the original geometry. While \sg can often appear like it tiles the geometry, the internal interactions of the subset do not add up to the original Hamiltonian.  Instead, all terms of the Hamiltonian are already in a single symmetry-reduced Hamiltonian.

%%%%%%%%%%%%%%%%%%%%%%%%%%%%%%%%%%%%%%%%%%%%%%%%%%%%%%%%%%%%%%
%%%%%%%%%%%%%%%%%%%%%%%%%%%%%%%%%%%%%%%%%%%%%%%%%%%%%%%%%%%%%%

\section{\sg in the Transverse Field Ising Model}\label{sec3}

\begin{figure}[b]
    \centering
    \includegraphics[width=0.98\columnwidth]{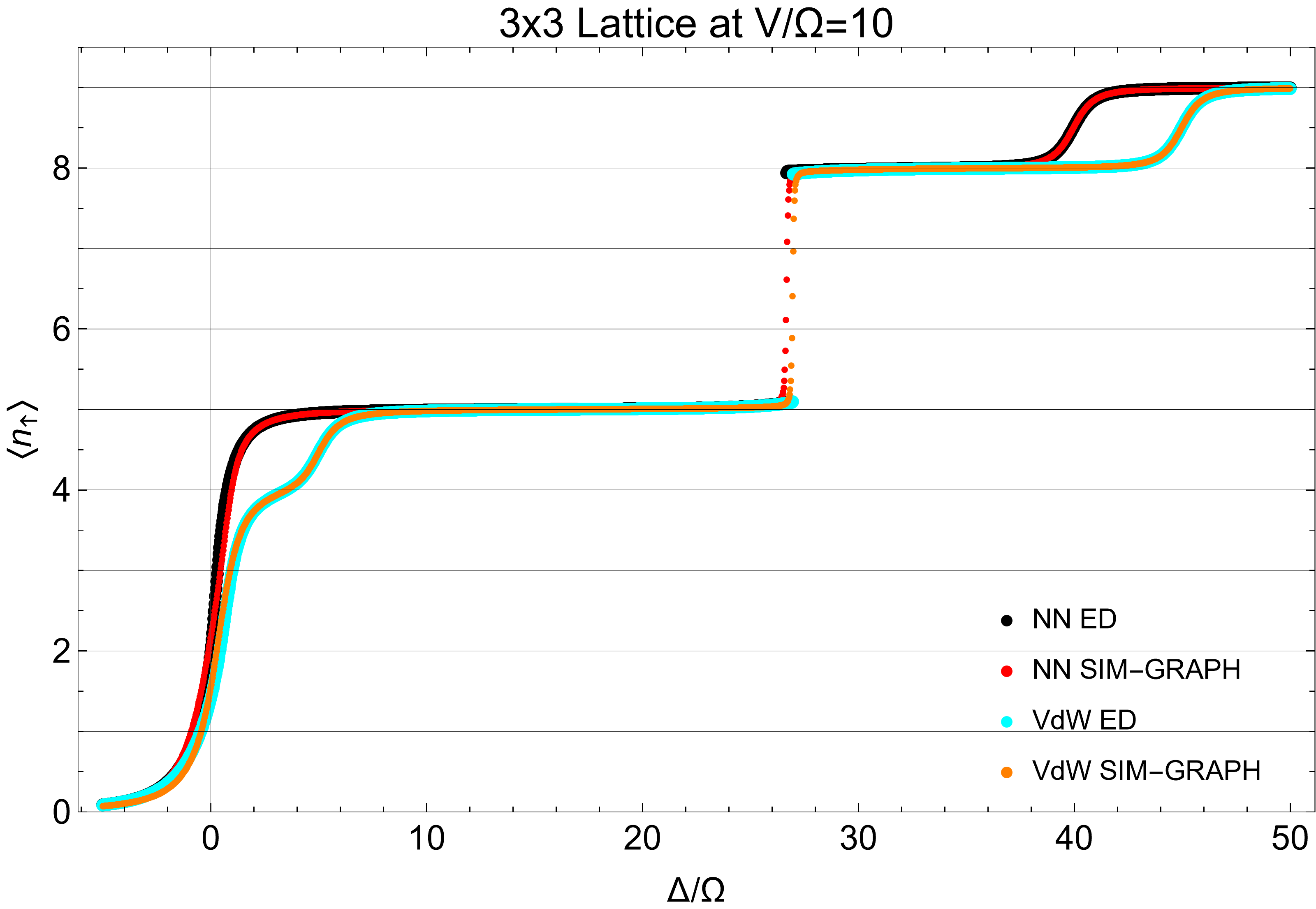}
    \caption{The average ground-state magnetization of the TFIM on the $3\times 3$ square lattice is presented at $V/\Omega=10$ for both NN and VdW interactions, computed using both ED and SIM-GRAPH. For better visual comparison, the exact results are plotted using larger points, with the SIM-GRAPH results overlaid in points half the size. For both NN and VdW interactions, we find excellent agreement with only minor deviations from the exact results. }
    \label{fig:3x3plot}
\end{figure}

The first application of \sg has been in the TFIM, which is often used in quantum simulators~\cite{verstraten2025control}. In this model, we begin with a spin 1/2 particle on every site. Then, we find three types of terms. First, there is a repulsive interaction $V_{ij}$ between nearby spin-up sites. Here, $V_{ij}$ could in principle be any type of interaction potential, but for the purposes of this article we will consider two distinct types: A short-range constant strength $V$ between each Nearest Neighbor (NN) and zero elsewhere, or a long-range Van der Waals (VdW) potential $V_{ij}= V (r_{ij}/a)^{-6}$, where $a$ is the NN distance. Second, there is a transverse field of strength $\Omega$ in the $x$ direction, which effectively mixes up and down states. Finally, we find an on-site detuning $\Delta$, which acts as a local potential that favors the spin-up state. The complete TFIM Hamiltonian is thus given by
\begin{equation}
H = \sum_{i<j} V_{ij}\, n_i^{z} n_j^{z}
\;+\;
\sum_{i=1}^{N} \left( \frac{ \Omega}{2}\,\sigma_i^{x} -  \Delta\, n_i^{z} \right).
\end{equation}

Let us consider how a symmetry operator $\hat{S}$ acts on this Hamiltonian. Take the same $3\times 3$ square from Fig.~\ref{fig:squarereduction} as an example, where the square is located within the $xy$-plane, with the central site 4 at the origin. Since the physical rotation does not alter the spin-space, we can quickly identify that $\hat{S}$ must simply relabel the indices according to either a rotation or a reflection. Thus, the Hamiltonian indeed obeys the same symmetry properties as discussed previously. Now, let us analyze the results generated by SIM-GRAPH and compare them to ED. 

\begin{figure}[t]
    \centering
    \includegraphics[width=0.98\columnwidth]{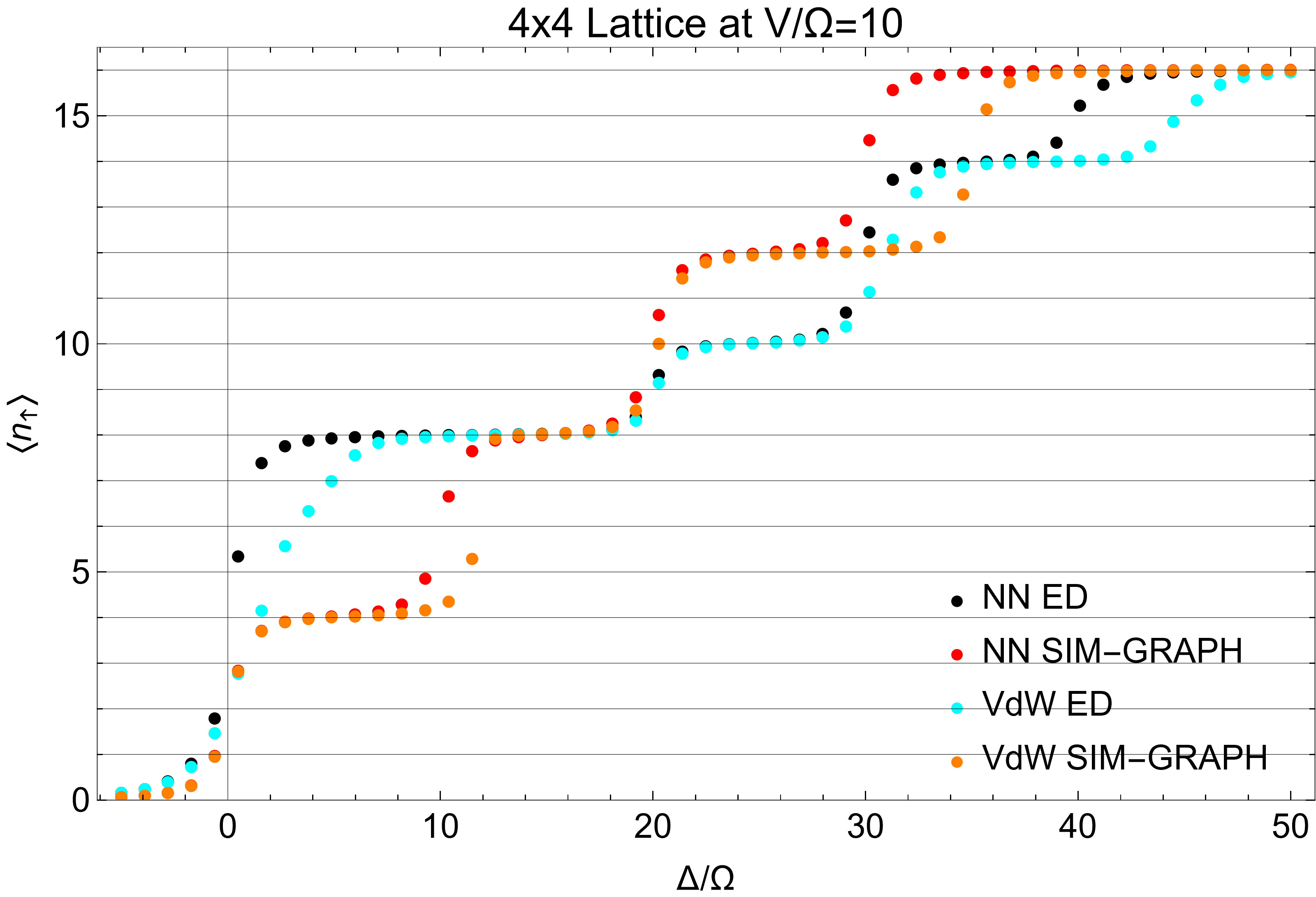}
    \caption{The average ground-state magnetization of the TFIM on the $4\times 4$ square lattice is presented at $V/\Omega=10$ for both NN and VdW interactions, computed on 50 points using both ED and SIM-GRAPH. For both NN and VdW interactions, we find deviations up to 4 spins away from the exact results. Besides the high and low $\Delta/\Omega$ limits, we find only one region of agreement at $\langle n_\uparrow \rangle=8$, which corresponds to the checkerboard states. }
    \label{fig:4x4plot}
\end{figure}

In Fig.~\ref{fig:3x3plot}, we show the ground-state magnetization for both short- and long-range interactions, computed using both ED and SIM-GRAPH. At negative detuning, the system has little incentive to have spins up, but when detuning is increased, it slowly ramps up to $\langle n_\uparrow \rangle=5$, which corresponds to a checkerboard pattern with all corners and the central sites excited. The long-range interactions cause an intermediate plateau, where the central site is still suppressed by the next-nearest-neighbor (NNN) interactions from the corners. When detuning is further increased, we find a big jump from 5 to 8 spins, where only the central site remains down, and all outer sites are excited. This state remains for a larger interval of $\Delta/\Omega$ in the case of long-range interactions, since it is suppressed by more --- and thus effectively stronger --- interactions. Finally, when the detuning is the dominant term, all 9 spins are in the up state.

In Fig.~\ref{fig:4x4plot}, we show the same results as before, but now for the even-sized $4\times 4$ square. Note that the symmetry-reduced graph contains the same three points as the $3\times 3$ square, although with different connections. In this case, we find much worse agreement than with the $3\times 3$ square. The predominant reason for the differences can be found by considering the half-excited checkerboard state at $\langle n_\uparrow \rangle=8$. Here, the exact ground state consists of a superposition of the two distinct checkerboard configurations. However, this state cannot be reproduced from the symmetric subset, since \sg assumes that all 4 inner sites must be in the same up or down state. In reality, it is energetically much more efficient to form a superposition of excited pairs across the two diagonals, but this option is projected out of the possible states by \sg.

%%%%%%%%%%%%%%%%%%%%%%%%%%%%%%%%%%%%%%%%%%%%%%%%%%%%%%%%%%%%%%%%
%%%%%%%%%%%%%%%%%%%%%%%%%%%%%%%%%%%%%%%%%%%%%%%%%%%%%%%%%%%%%%%%

\section{\sg 2: Partial Reductions}\label{sec4}

\begin{figure}[b]
    \centering
    \includegraphics[width=0.95\columnwidth]{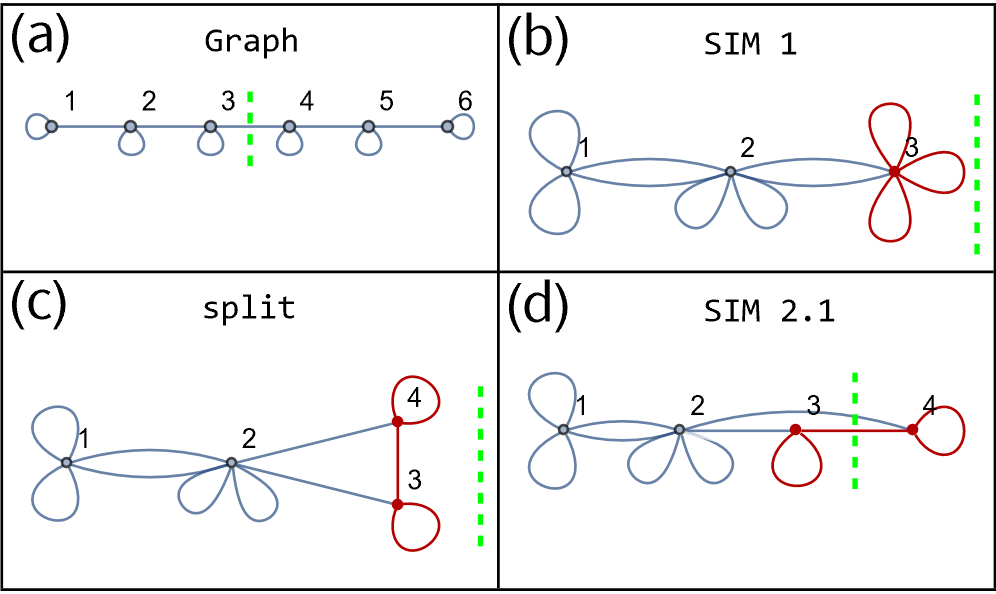}
    \caption{(a) The NN interaction graph for a chain of 6 spins. The dashed green line indicates the symmetry axis. (b) The reduced graph based on SIM1. (c) Opening of the 3 to 4 link, which allows for a superposition across the symmetry axis. (d) The reduced graph based on SIM2.1. In (b)-(d), the red highlight indicates the difference between SIM1 and SIM2.1.}
    \label{fig:chain6}
\end{figure}

We will refer to the complete symmetry projection of the previous sections as SIM1. Here, we will present an extension of SIM1, where this projection is only partially applied. This method, which we will refer to as SIM2, is objectively slower than SIM1. It may also be less accurate when symmetry is not taken into account with care. Nevertheless, when used appropriately, SIM2 may predict correct ground-state properties for a much wider range of geometries than SIM1. The most important reason that SIM2 may predict better than SIM1, is that it may capture superpositions across the symmetry axis. Since correlations often tend to decay with distance~\cite{hastings2006spectral}, these potentially missed superpositions are most likely found close to the symmetry axis. Thus, one might consider to retain these points in the reduction procedure. For this extension to work effectively, it is crucial that the non-projected site(s) retain the internal symmetry with their symmetric representation(s). Otherwise, the two sites could wrongly predict different observables.

We will employ the nomenclature SIM2.$n$, where $n=0,1,2,\dots$ includes all neighbors up to the $n^{th}$ NN of the symmetric sites that are retained in the projection. Note that SIM2.0 is thus equivalent to SIM1. As an example, we consider Fig.~\ref{fig:chain6}(a), where the interaction graph of 6 NN interacting spins is presented. We note that the symmetry axis (dashed green) cuts the chain between sites 3 and 4. The symmetry projection is given by $P_{S}: (1\mapsto 1,2\mapsto 2,3\mapsto 3,4\mapsto 3,5\mapsto 2,6\mapsto 1)$. Thus, in SIM1 this chain is projected to the graph in Fig.~\ref{fig:chain6}(b), where we see that site 4 is projected onto site 3 (see red highlight). This means that there is no possibility for a superposition across the original symmetry axis, even though this is a likely contribution to the ground state. This issue is resolved by untangling sites 3 and 4, see red highlight in Fig.~\ref{fig:chain6}(c). Here, we can also clearly see that sites 3 and 4 remain fully symmetric in this partially reduced graph. In Fig.~\ref{fig:chain6}(d), we have simply repositioned the vertices to the original locations of the full length chain, which does not alter the internal symmetry. This partially reduced graph is also directly produced by applying SIM2.1 to the full length chain using the projection operator $P_{S2,1}: (1\mapsto 1,2\mapsto 2,3\mapsto 3,4\mapsto 4,5\mapsto 2,6\mapsto 1)$.

% \begin{figure*}[ht]
%     \centering
%     \includegraphics[scale=0.47]{fig/fig1.png}
%     \caption{}
%     \label{fig:FE}
% \end{figure*}

\section{\sg 3: Partial Symmetries}\label{sec5}

Here, we present a slight variation on the idea of SIM2, where we do not consider partial reductions of the graph, but subgroups of the symmetry class. In most cases that we have considered, there is both a rotational and a mirror symmetry. When the ground state breaks the symmetry of the system, it will often only break one of these two symmetries. Thus, by considering the reduced model on the other symmetry, we retain the capability to find the correct symmetry-broken state. For example, the ground state of a triangle might consist of a left-handed plus right-handed symmetry, which both break the mirror symmetry, while retaining rotational symmetry. In such a case, considering the graph reduction only based on the rotational symmetry will allow to compute the correct states. The procedure is identical to SIM1, except that we manually choose a subgroup of the symmetry class to use for the projection. We will call this method SIM3, and we will by default choose the rotational symmetry for reduction, unless stated otherwise. Remark that, for some geometries, this change might not make any difference in the outcome, as it is the case for the 3x3 square in Fig.~\ref{fig:squarereduction}, where SIM1 and SIM3 both produce the same subgraph. However, for the 4x4 square lattice there is a difference, since the rotational symmetry distinguishes the two NN of the corner. In certain cases, SIM3 might work much better than SIM2, depending on which correlations are most important in the geometry, but the reverse may also be true.

%%%%%%%%%%%%%%%%%%%%%%%%%%%%%%%%%%%%%%%%%%%%%%%%%%%%%%%%%%%%%%
%%%%%%%%%%%%%%%%%%%%%%%%%%%%%%%%%%%%%%%%%%%%%%%%%%%%%%%%%%%%%%

\section{Toy model examples of \sg}\label{sec6}
In this section, we explicitly recall the principles of \sg and apply the method to some minimal working models. This section is separated into two parts; first we consider odd and then even symmetries. These two symmetries are defined by the place where the symmetry axis divides the sites, either cutting through them (odd) or cutting between them (even). We will see that this distinction is important due to possible superpositions across the symmetry axis.

\subsection{Odd symmetries: A chain of 3 NN interacting spin-1/2 particles  (SIM1)}
Consider a minimal working model: three spin-1/2 particles ($\hbar=1$) interacting with strength $V$, in a transverse field $\Omega$ and a longitudinal field $\Delta$, given by
\begin{equation}
H = \sum_{\langle i,j\rangle} V\, n_i^{z} n_j^{z}
\;+\;
\sum_{i=1}^{3} \left( \frac{ \Omega}{2}\,\sigma_i^{x} -  \Delta\, n_i^{z} \right),
\end{equation}
or in matrix form
\begin{equation}
 \fiteq{H_3=  \begin{pmatrix}
\ket{000} & \ket{001} & \ket{010} & \ket{011} & \ket{100} & \ket{101} &\ket{110} & \ket{111} \\
0 & \Omega/2 & \Omega/2 & 0 & \Omega/2 & 0 & 0 & 0 \\
\Omega/2 & -\Delta & 0 & \Omega/2 & 0 & \Omega/2 & 0 & 0 \\
\Omega/2 & 0 & -\Delta & \Omega/2 & 0 & 0 & \Omega/2 & 0 \\
0 & \Omega/2 & \Omega/2 & V - 2\Delta & 0 & 0 & 0 & \Omega/2 \\
\Omega/2 & 0 & 0 & 0 & -\Delta & \Omega/2 & \Omega/2 & 0 \\
0 & \Omega/2 & 0 & 0 & \Omega/2 & -2\Delta & 0 & \Omega/2 \\
0 & 0 & \Omega/2 & 0 & \Omega/2 & 0 & V - 2\Delta & \Omega/2 \\
0 & 0 & 0 & \Omega/2 & 0 & \Omega/2 & \Omega/2 & 2V - 3\Delta
\end{pmatrix}. }
\end{equation}
Note that this system has a mirror symmetry, which can be described by the cycles $\{(1,3),(2)\}$, where the two ends are exchangeable.

\subsubsection{Matrix Reduction}
The symmetry of this system implies that any observable should have equal value on both ends, site 1 and 3. More explicitly, we have a symmetry operator $S: (1,2,3) \mapsto (3,2,1)$, which commutes with $H$. Thus, $H$ and $S$ have the same eigen-spaces. Since $S^2=\mathbb{1}$, its eigenvalues are $\lambda=\pm1$. This allows us to block-diagonalize $H$ as 
\begin{equation}
    H=\begin{pmatrix}
        H_+ &0\\ 0 &H_-
    \end{pmatrix}.
\end{equation}
Let $\ket{\psi}=\{a_1,\dots, a_8\}$ be the ground state in the binary basis, $$B_3= \{000,001,010,011,100,101,110,111\},$$ then $H(S\ket{\psi})=S(H\ket{\psi})=S(E\ket{\psi})=E(S\ket{\psi})$ implies that $\ket{\psi}$ is an eigenvector of $S$, and must lie in either the + or the - sector of $H$. In fact, if even one of the symmetric components is non-zero, then we know that its symmetry eigenvalue must be +1,  $S\ket{\psi}=\ket{\psi}$, and thus $a_2=a_5$ and $a_4=a_7$. This assumption narrows down the Hilbert space $H$ from 8 to 6 dimensions in $H_+$. However, this still leaves us with two degrees of freedom in non-symmetric states, namely $\ket{100}_+ =(\ket{001}+\ket{100})/\sqrt{2}$ and $\ket{110}_+ =(\ket{011}+\ket{110})/\sqrt{2}$. Let us divide $H_+$ into a singlet and a doublet state sector
\begin{align}
    &H_+= \begin{pmatrix}
        H_{11} &H_{12}\\ H_{21} & H_{22}
    \end{pmatrix}   \\
    &=
    \left(
\begin{array}{cccc|cc}
\ket{000} & \ket{010} & \ket{101} & \ket{111} & \ket{100}_+ & \ket{110}_+   \\
0 & \Omega/2 & 0 & 0 & \Omega/\sqrt{2} & 0 \\
\Omega/2 & -\Delta & 0 & 0 & 0 & \Omega/\sqrt{2} \\
0 & 0 & -2\Delta & \Omega/2 & \Omega/\sqrt{2} & 0 \\
0 & 0 & \Omega/2 &  2V-3\Delta  & 0 & \Omega/\sqrt{2} \\
\hline
\Omega/\sqrt{2} & 0 & \Omega/\sqrt{2} & 0 & -\Delta & \Omega/2 \\
0 & \Omega/\sqrt{2} & 0 & \Omega/\sqrt{2} & \Omega/2 & V - 2\Delta
\end{array}
% \begin{array}{cccc|cc}
% \ket{000} & \ket{010} & \ket{101} & \ket{111} & \ket{100}_+ & \ket{110}_+   \\
% 0 & \Omega/2 & 0 & 0 & \Omega/2 & 0 \\
% \Omega/2 & -\Delta & 0 & 0 & 0 & \Omega/2 \\
% 0 & 0 & -2\Delta & \Omega/2 & \Omega/2 & 0 \\
% 0 & 0 & \Omega/2 & -3\Delta + 2V & 0 & \Omega/2 \\
% \hline
% \Omega/2 & 0 & \Omega/2 & 0 & -\Delta & \Omega/2 \\
% 0 & \Omega/2 & 0 & \Omega/2 & \Omega/2 & V - 2\Delta
% \end{array}
\right). \nonumber
\end{align}
% Next, we may perform an isospectral reduction to the singlet sector, which is how we encode the doublet states into effective singlet states.
% \begin{equation}
%     H_{eff}[E]= H_{11} + H_{12}\frac{1}{E-H_{22}}H_{21},
% \end{equation}
% where the effective Hamiltonian depends on the energy $E$.

Finally, our aim is to only consider the singlet symmetric sector. However, we need to investigate the effect of the transverse field on the doublets:
\begin{align}
    \sum_i \frac{\Omega}{\sqrt{2}} \sigma_i^x \ket{100}_+ &= \Omega(\ket{000} +\ket{101}) +\frac{\Omega}{2} \ket{110}_+;\\
     \sum_i \frac{\Omega}{\sqrt{2}} \sigma_i^x \ket{110}_+ &= \Omega(\ket{010} +\ket{111}) +\frac{\Omega}{2} \ket{100}_+.
\end{align}
Thus, when projecting out the doublet states, we get new effective transitions of strength $\Omega$ between certain singlet states, mediated via the doublet state. We may also interpret this from the graph perspective. Here, the transverse field terms from site 3 are projected to site 1, causing a twice as strong local magnetic field on site 1 (as compared to site 2), which is now responsible for a double flip in the total spin count. Adding these effective terms to the $H_{11}$ sector, we finally obtain the \sg effective Hamiltonian
\begin{equation}
    H_{SIM}=
    \begin{pmatrix}
       \ket{000} & \ket{010} & \ket{101} & \ket{111}\\
0 & \Omega/2 & \Omega & 0 \\
\Omega/2 & -\Delta & 0 & \Omega \\
\Omega & 0 & -2\Delta & \Omega/2 \\
0 & \Omega & \Omega/2 & 2V - 3\Delta
\end{pmatrix}.\label{eq:sim1}
\end{equation}

\begin{figure}[t]
    \centering
    \includegraphics[width=0.95\columnwidth]{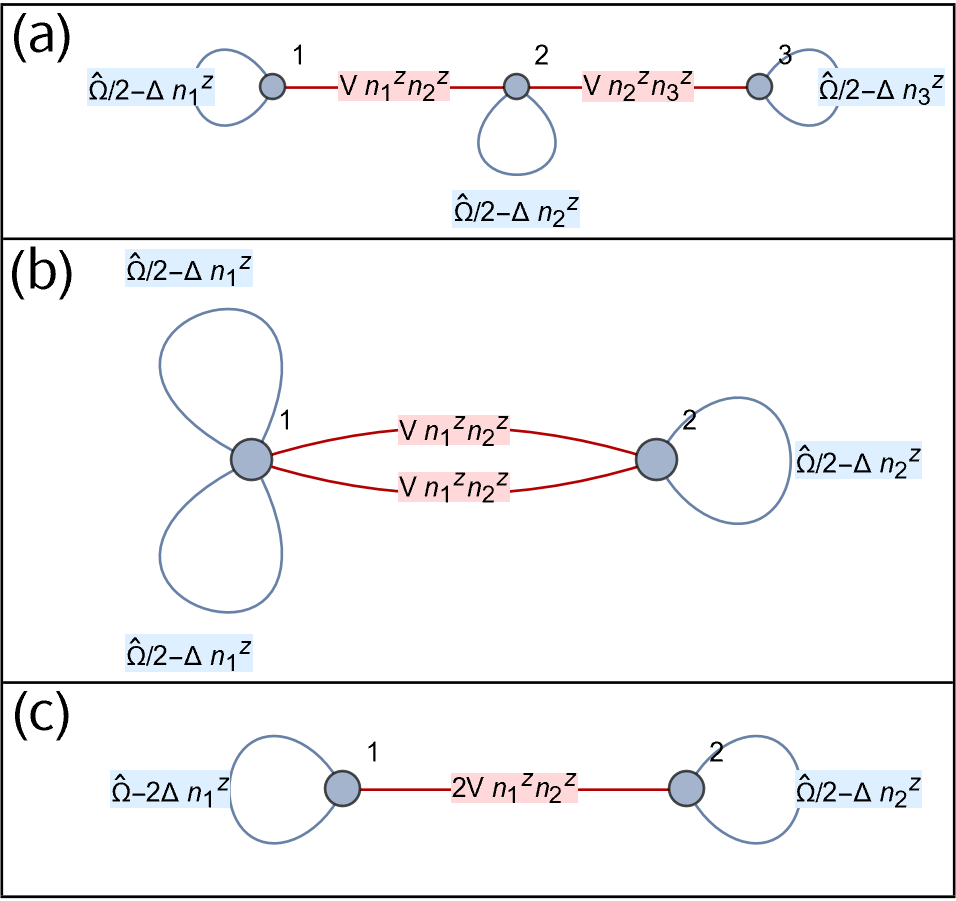}
    \caption{The NN interaction graph for a chain of 3 spins is shown and projected according to the \sg procedure. Here, $\hat{\Omega}= \Omega\sum_i \sigma_i^x$. (a) The original graph of interactions in the NN Ising chain of 3 sites. (b) The terms connected to site 3 have been projected to site 1. (c) Double terms are gathered and the symmetry reduced graph is produced.}
    \label{fig:chain3}
\end{figure}

\subsubsection{Graph Reduction}
Let us now consider the same calculation directly from the graph perspective, see Fig.~\ref{fig:chain3}. Firstly, we place all terms of the Hamiltonian onto the interaction graph, see Fig.~\ref{fig:chain3}(a), where local terms (blue) are stored in loops connected to a single point and non-local terms (red) are links between different sites. Next, sites 1 and 3 are symmetric, which means that we will project all terms from site 3 onto site 1. This produces the graph presented in Fig.~\ref{fig:chain3}(b). Double connections can be simplified by adding the terms to a single connection, revealing the final symmetry reduced graph in Fig.~\ref{fig:chain3}(c). We can convert this graph back to the Hamiltonian matrix, for which we now use the two digit binary basis, leading to
\begin{equation}
    H_{3,SIM}=
    \begin{pmatrix}
        \ket{00} & \ket{01} & \ket{10} & \ket{11 }\\
0 & \Omega/2 & \Omega & 0 \\
\Omega/2 & -\Delta & 0 & \Omega \\
\Omega & 0 & -2\Delta & \Omega/2 \\
0 & \Omega & \Omega/2 & 2V - 3\Delta
\end{pmatrix}.
\end{equation}
Note that this is indeed identical to the Hamiltonian derived in Eq.~\eqref{eq:sim1}, where we find the third spin state by copying the state on site 1 to site 3. Furthermore, remark how little computation has been required to derive this effective Hamiltonian. Once symmetry relations between sites are established, all remaining tasks are basic additions. There is never any consideration of singlets vs doublets, block-diagonalization, or effective transition rates.

\subsection{Even symmetries: A chain of 4 NN interacting spin-1/2 particles (SIM2)}
Next, we consider the minimal working model with an even symmetry. We will demonstrate what is lacking in the previous method, and how minor adjustments may solve these issues. To begin, let us directly apply the \sg principles to the graph of interactions. We obtain a result very similar to Fig.~\ref{fig:chain3}(c), but with slightly different weights. This defines the Hamiltonian
\begin{equation}
       \tilde{H}_{4}=
    \begin{pmatrix}
        \ket{00} & \ket{01} & \ket{10} & \ket{11 }\\
0 & \Omega & \Omega & 0 \\
\Omega & V-2\Delta & 0 & \Omega \\
\Omega & 0 & -2\Delta & \Omega \\
0 & \Omega & \Omega & 3V - 4\Delta
\end{pmatrix}. 
\end{equation}
Now, considering that $V,\Delta >0$, we should notice that the lowest energy basis states have energy of either $-2\Delta$, or $3V-4\Delta$ if $\Delta>3V/2$. However, notice that the states $\ket{1010}$ and $\ket{0101}$ also have energy $-2\Delta$, but are excluded from these symmetric options. Thus, we are likely to miss the ground state in the intermediate $\Delta$ regime. The solution to capture these states is to include sites which lie just beyond the symmetry axis, in this case site 3. Thus, for this 4 site chain, we would only project site 4 onto site 1. This yields the effective Hamiltonian
\begin{align}
&  H_{4,SIM}= \\ 
 & \fiteq{ \begin{pmatrix}
\ket{000} & \ket{001} & \ket{010} & \ket{011} & \ket{100} & \ket{101} &\ket{110} & \ket{111} \\
0 & \Omega/2 & \Omega/2 & 0 & \Omega & 0 & 0 & 0 \\
\Omega/2 & -\Delta & 0 & \Omega/2 & 0 & \Omega & 0 & 0 \\
\Omega/2 & 0 & -\Delta & \Omega/2 & 0 & 0 & \Omega & 0 \\
0 & \Omega/2 & \Omega/2 & V - 2\Delta & 0 & 0 & 0 & \Omega \\
\Omega & 0 & 0 & 0 & -2\Delta & \Omega/2 & \Omega/2 & 0 \\
0 & \Omega & 0 & 0 & \Omega/2 & V-3\Delta & 0 & \Omega/2 \\
0 & 0 & \Omega & 0 & \Omega/2 & 0 & V - 3\Delta & \Omega/2 \\
0 & 0 & 0 & \Omega & 0 & \Omega/2 & \Omega/2 & 3V - 4\Delta
\end{pmatrix}. }\nonumber
\end{align}
Notice how this new Hamiltonian is still symmetric under exchange of sites 2 and 3.

\begin{figure}[b]
    \centering
    \includegraphics[width=\columnwidth]{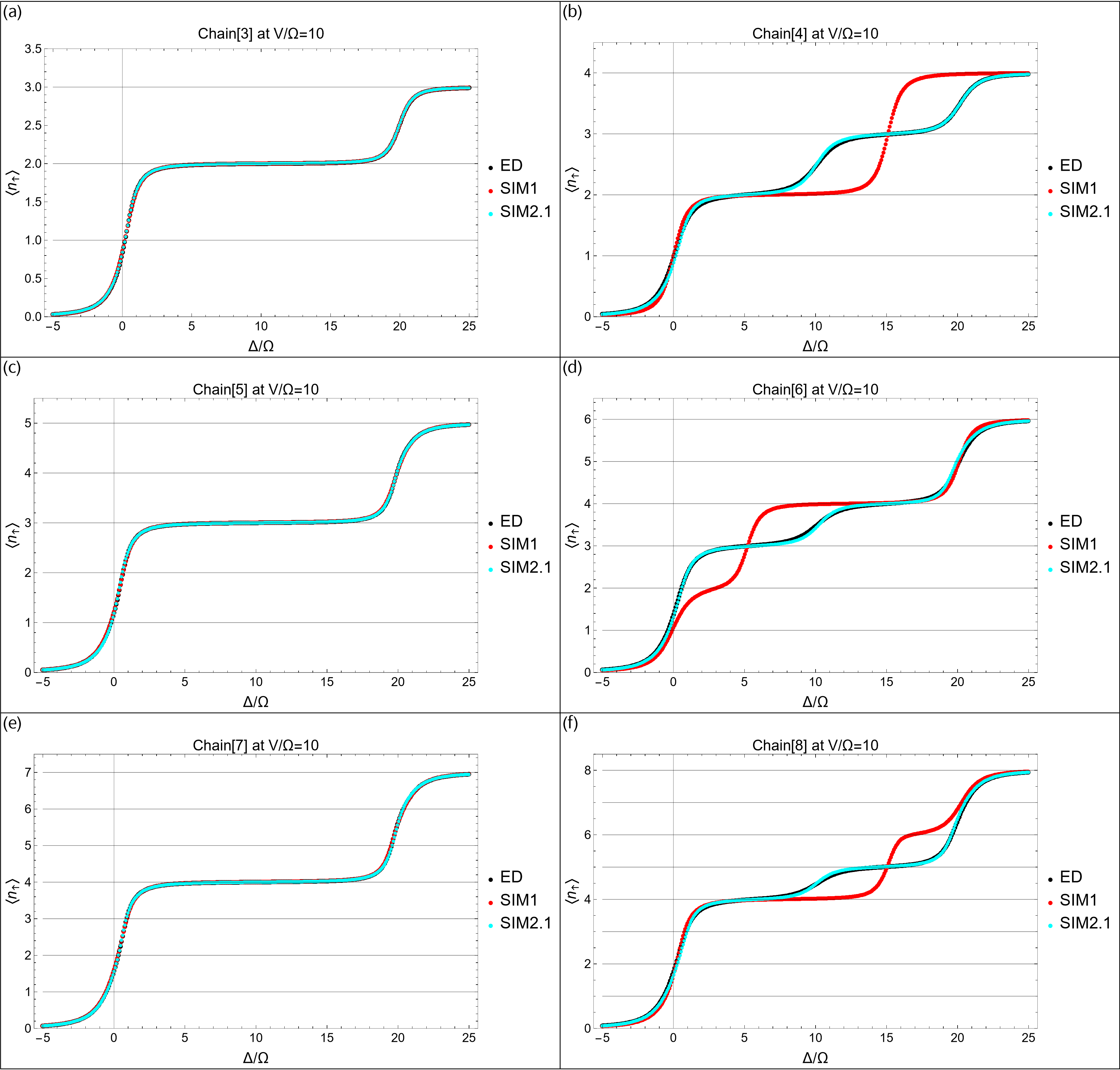}
    \caption{The number of spin-up excitations in the groundstate $ \langle n_\uparrow \rangle $ is compared for chains between length (a) 3 to (f) 8 with VdW interactions. The exact results are in black, SIM1 in red, and SIM2.1 in cyan. SIM1 and SIM2.1 yield similarly good results for odd lengths (left column), but SIM2.1 is far superior for even lengths (right column).}
    \label{fig:chaincomparison}
\end{figure}

Let us now consider a specific example with $\Omega=1, V=10, \Delta=13$, chosen such that we are in a regime where SIM1 fails, see Fig.~\ref{fig:chaincomparison}(b). Here, the exact ground state is approximately $\ket{\psi_0}= 0.68\ket{1011} + 0.68\ket{1101}+ \mathcal{O}$, where other states contain less than 5 percent probability, and an eigenenergy of approximately $E_0=-29.3 \approx V-3\Delta$. In SIM1, the ground state consists of $\ket{\psi_1}=-0.97\ket{1001}+0.23\ket{1111} + \mathcal{O}$ with eigenenergy $E_1=-26.3 \approx -2\Delta$. In SIM2, we find that $\ket{\psi_2}= 0.68\ket{1011}+0.68\ket{1101} + \mathcal{O}$ with eigenenergy $E_2=-29.3 \approx V-3\Delta$, just like the exact ground state. However, we must be careful not to overstate the accuracy of SIM2. Note, for example, all the slight deviations in the magnetization of ED vs SIM2 in Fig.~\ref{fig:chaincomparison}(b,d,f). These variations correspond precisely to the minor amplitudes in non-symmetric states which are neglected by SIM2, but are present in ED.

In fact, one can even go further because there are still regimes where the local magnetization from SIM2 is wrong. Consider, for example, $\Omega=1, V=10, \Delta=5$, which at first glance in Fig.~\ref{fig:chaincomparison}(b) seems to be a good prediction by both SIM1 and SIM2. However, while the average magnetization is correct, the local spin up probability from ED is $p_i=(0.77, 0.23, 0.23, 0.77)$, while SIM1 predicts $p_i=(0.99, 0.00, 0.00, 0.99)$ and SIM2 predicts $p_i=(0.99, 0.01, 0.01, 0.99)$. Here, \sg has found part of the ground state $\ket{1001}$, but has missed two degenerate states $\ket{1010}$ and $\ket{0101}$, which both have the same eigenenergy $-2\Delta$. This can be solved in two ways. Firstly, long-range interactions will provide a small energy gap between these states, favoring those with spins further away from each other, which is precisely the symmetric state captured by \sg. The other way to capture these extra states is to include them in the reduction, extending the chain by an extra site. While this is fruitless for this short chain, for larger systems we can extend the accuracy of \sg, by including either NN of the symmetric set, or go beyond and include NNN, or more. Even though this costs a significant amount of computational time, the speedup still scales exponentially faster in the total system size.

%%%%%%%%%%%%%%%%%%%%%%%%%%%%%%%%%%%%%%%%%%%%%%%%%%%%
%%%%%%%%%%%%%%%%%%%%%%%%%%%%%%%%%%%%%%%%%%%%%%%%%%%%%%%
%new

\section{Results: A \sg comparison for different geometries}\label{sec7}

%Now, we compare some results between the different variations of \sg. 
\subsection{One-Dimensional Chains}
In Fig.~\ref{fig:chaincomparison}, we show the total ground state magnetization $ \langle n_\uparrow \rangle $ for several one-dimensional chains, ranging from 3 to 8 sites in length. We see that all methods yield nearly identical results for odd-length chains. For even-length chains, however, we observe a large discrepancy between ED and SIM1 in the intermediate regime. Here, the dominant contributions are formed from anti-symmetric pairs, which are missed by the projection of SIM1. SIM2.1 allows for the superposition of the anti-symmetric pair across the symmetry axis, which permits to simulate one-half of this anti-symmetric state. By the imposed symmetry, this is enough information to construct the local and average magnetization, and we find that this indeed matches excellently with ED.

\begin{figure}[b]
    \centering
    \includegraphics[width=0.98\columnwidth]{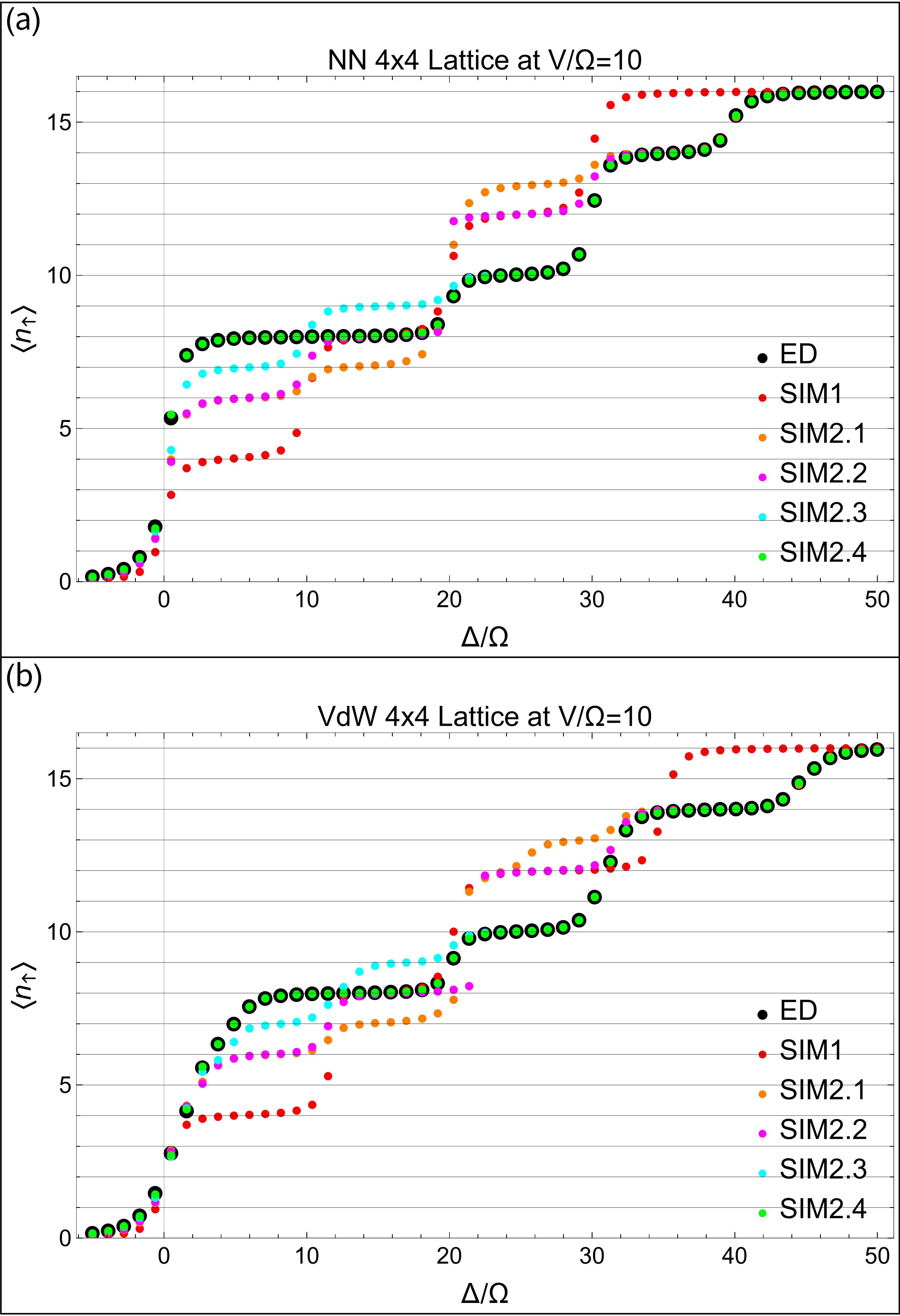}
    \caption{The convergence of SIM2.n is presented for the $4\times 4$ square lattice TFIM with (a) NN  and (b) VdW interactions. The number of used sites are 3 for SIM1, 7 for SIM2.1, 9 for SIM2.2, 12 for SIM2.3, 15 for SIM2.4, and 16 for ED. We can observe a clear convergence upon increasing the number of retained neighbors. In both cases, we see that SIM2.4 no longer shows any significant deviations from ED, despite the fact that it has only half the Hilbert space dimension as compared to ED. Finally, remark that none of the methods ever deviate by more than 4 spins from ED, which is precisely the order of the square-lattice symmetry.}
    \label{fig:4x4sim2}
\end{figure}

\subsection{Two-Dimensional $4 \times 4$ Square}
Let us now consider the case for which we demonstrated SIM1 to fail significantly, namely the $4\times 4$ square lattice. In Fig.~\ref{fig:4x4sim2}, we compare the magnetization of SIM2.n for (a) NN and (b) VdW interactions to the exact results. We see that it takes many more neighbors to remove all deviations, but there is a clear convergence to the exact results. This convergence is two-fold; both the size of the deviations, as well as the region of deviations are reduced. One may wonder why this geometry requires so many more extra neighbors in order to get matching results, as compared to the one-dimensional chain, which only required one extra neighbor. The reason is that there are many more likely superpositions in the two-dimensional case, since there are more neighbors next to each site. The specific geometry also plays a role. Consider the four central sites, for instance. In SIM1, these are all assumed to be identical, but in reality there are many partially filled configurations which are very likely, such as a single spin in superposition across the four sites, or two spins in superposition across the two diagonals. These central asymmetries can also cause asymmetry in the outer sites, which in turn yields a highly entangled symmetric state consisting of many asymmetric parts. While a higher $n$ may reduce the number of missed asymmetries, these long-range superpositions are always the first to be projected out by \sg, and so will be a weakness of the method, costing much more computational time to resolve.

\subsection{Frustrated Sierpinski Triangle}

\begin{figure}[b]
    \centering
    \includegraphics[width=0.98\columnwidth]{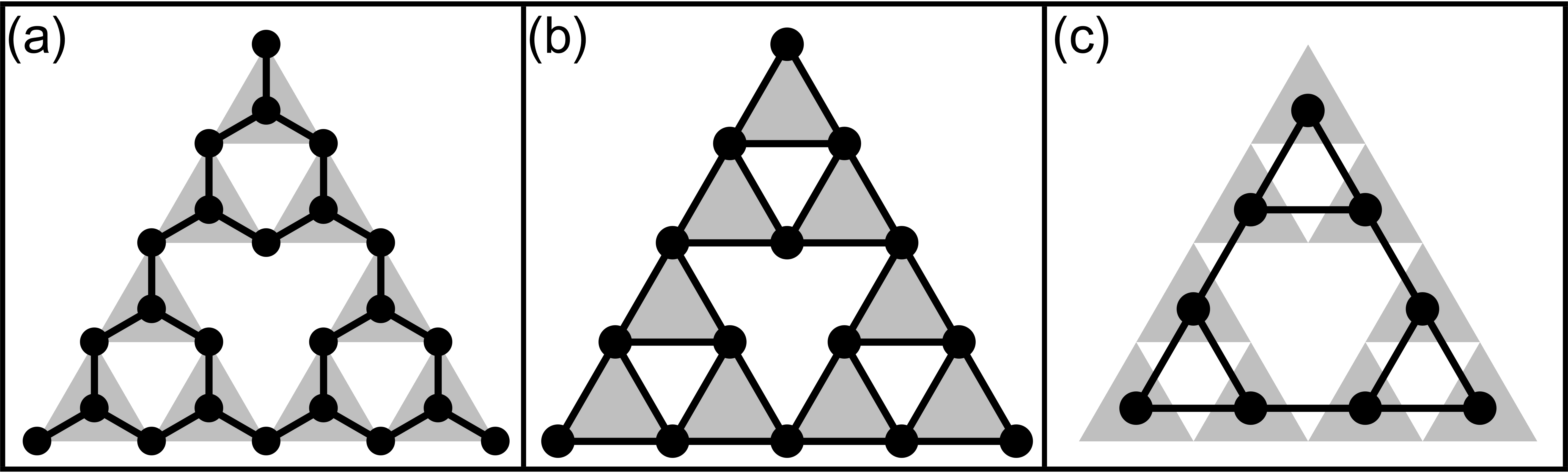}
    \caption{Three types of lattices that may be defined on the second-generation Sierpinski triangle, depicted by the gray areas. (a) The hexagonal lattice is constructed by placing sites both at the center and at the corners of each triangle. (b) The triangular lattice is formed by only placing sites on the corners of each triangle. (c) The dual triangular lattice is constructed by placing a single site at the center of each triangle. Note that this geometry misses the smallest hole, which is why we consider this geometry as effectively a first-generation triangle.}
    \label{fig:sierplat}
\end{figure}

\begin{figure}[t]
    \centering
    \includegraphics[width=0.98\columnwidth]{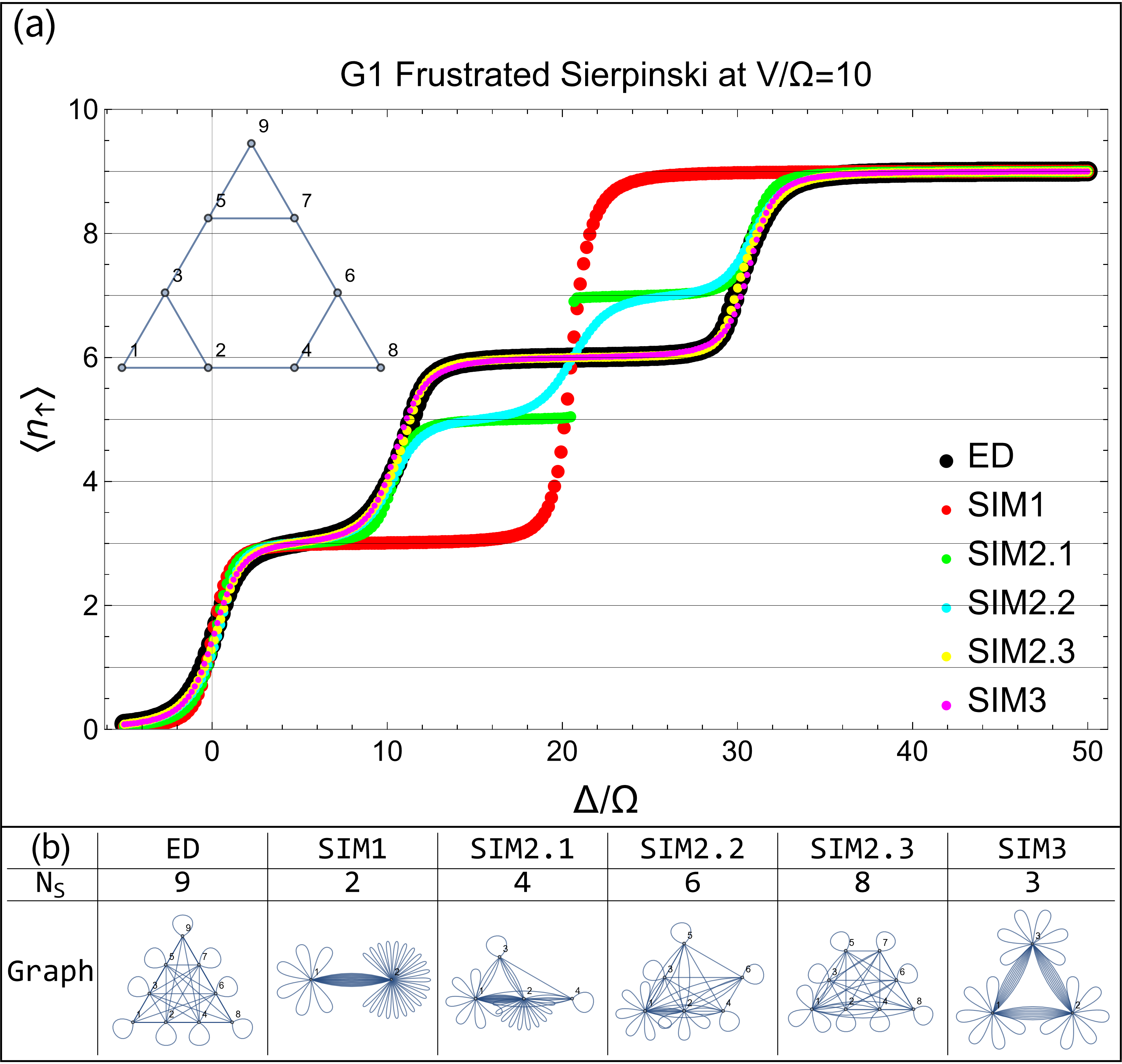}
    \caption{The various \sg methods are compared for the first-generation frustrated Sierpinski triangle with VdW interactions. (a) The ground-state magnetization is presented using the various \sg methods. (b) The effective graph used to calculate the magnetization for each method, with the number of effective sites $N_{S}$ indicated. For visual clarity, only NN bonds have been drawn in, but the results are computed using VdW interactions.  While we do find convergence of SIM2.n toward the exact results, we find that SIM3 immediately finds the correct groundstates. Meanwhile, SIM3 only uses 3 sites, whereas we needed 8 sites to get equally good results with SIM2.3  }
    \label{fig:sierpfrus}
\end{figure}

Next, we consider a geometry which strongly favors rotational symmetry, namely a frustrated triangle. In Fig.~\ref{fig:sierplat}, we show three ways to construct lattices on the Sierpinski triangle. To do this, one needs to define a connection between the continuous triangles and discrete points. In the hexagonal Sierpinski triangle, this is achieved by putting one point in the center and one point at each corner of the triangle [Fig.~\ref{fig:sierplat}(a)]. For the triangular lattice one only takes the corners of each triangle [Fig.~\ref{fig:sierplat}(b)], whereas for the dual lattice, one only takes points at the center of each triangle [Fig.~\ref{fig:sierplat}(c)].  The hexagonal substructure in Fig.~\ref{fig:sierplat}(a) is free of frustration due to the even number of connections around the holes of the lattice. However, the other two lattices have a triangular geometry, which adds frustration. The structure in Fig.~\ref{fig:sierplat}(b) requires many sites for capturing the geometry. Thus, let us consider the dual lattice depicted in Fig.~\ref{fig:sierplat}(c). Here, we should expect frustration within each small triangle. In Fig.~\ref{fig:sierpfrus}(a), we show the computed number of spin excitations in the groundstate for all of the \sg methods and we show the reduced graphs in Fig.~\ref{fig:sierpfrus}(b).  While we can see in Fig.~\ref{fig:sierpfrus}(a) that SIM2.n slowly converges to the exact solution, SIM3 is extremely accurate while using only 3 effective sites. That is because this geometry strongly favors mirror-broken states, allowing for 6 excitations without a NN interaction between the sub-triangles. This type of partial symmetry breaking is precisely the use-case where SIM3 works best.

\begin{figure}[b]
    \centering
    \includegraphics[width=0.98\columnwidth]{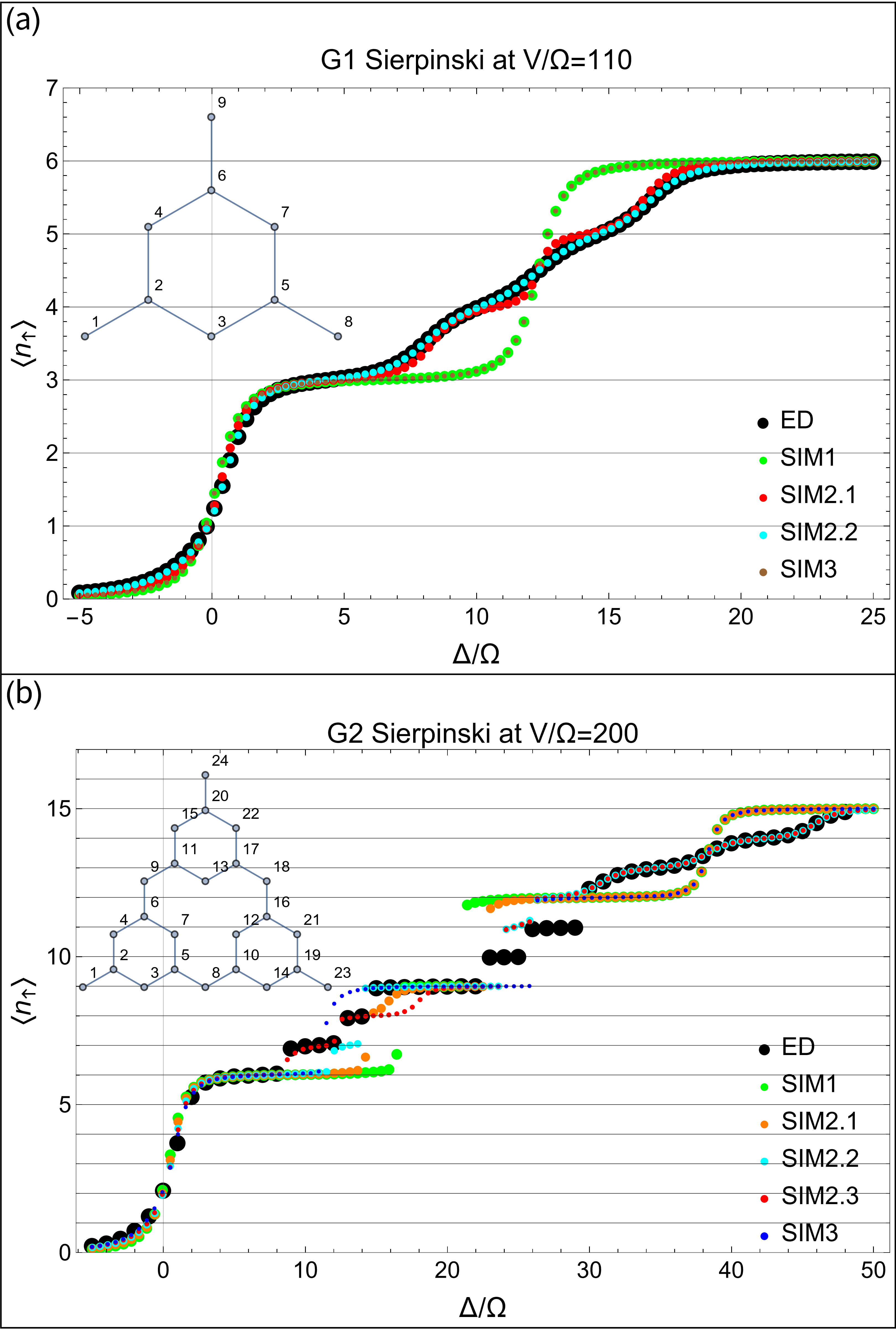}
    \caption{The convergence of SIM2.n and SIM3 is presented with VdW interactions on the first (a) and second (b) generation Sierpinski triangle. These are presented for the precise values that were also used in Ref.~\cite{verstraten2025control}. We observe a good convergence to the exact results using SIM2. Note, however, that there are still some errors in (b) for the largest subgraph presented, particularly around the $n=10$ and $n=11$ states. This is also the region where QMC had the most convergence issues because there are many configurations that are closely matched in energy. }
    \label{fig:sierpsim2}
\end{figure}

\subsection{Hexagonal Sierpinski Triangle}

Now, let us compare the performance of SIM2 on the geometries previously investigated in Ref.~\cite{verstraten2025control}, namely the first and second generation of the fractal Sierpinski triangle. These are the first two stages of a self-repeating structure which forms a fractal in the limit of infinite recursion. The Sierpinski triangle is recursively generated by taking a triangle and copying it three times to form a bigger triangle with a hole in the center. These geometries have the remarkable property that their dimension is non-integer, which can have profound consequences on the physics. In Fig.~\ref{fig:sierpsim2}, we compare the results from Ref.~\cite{verstraten2025control} (ED and SIM1) to the results computed using SIM2. We find that SIM2.n performs better and better the larger n becomes. For the first generation, Fig.~\ref{fig:sierpsim2}(a), we see that SIM2.1 already captures all features of the exact results, whereas SIM2.2 has nearly perfect agreement. For the second generation, Fig.~\ref{fig:sierpsim2}(b), the results mostly converge towards the exact results, with only a few regions containing significant errors in SIM2.3 as compared to ED. Nevertheless, computational time has increased compared to SIM1. On a laptop, the 100 points for G2 took between $(0.18 s,\; 0.96 s,\; 1min\; 31s,\;5min\;5s) $ to compute for SIM2.n $=0\dots 3$, respectively.

\begin{figure}[b]
    \centering
    \includegraphics[width=0.98\columnwidth]{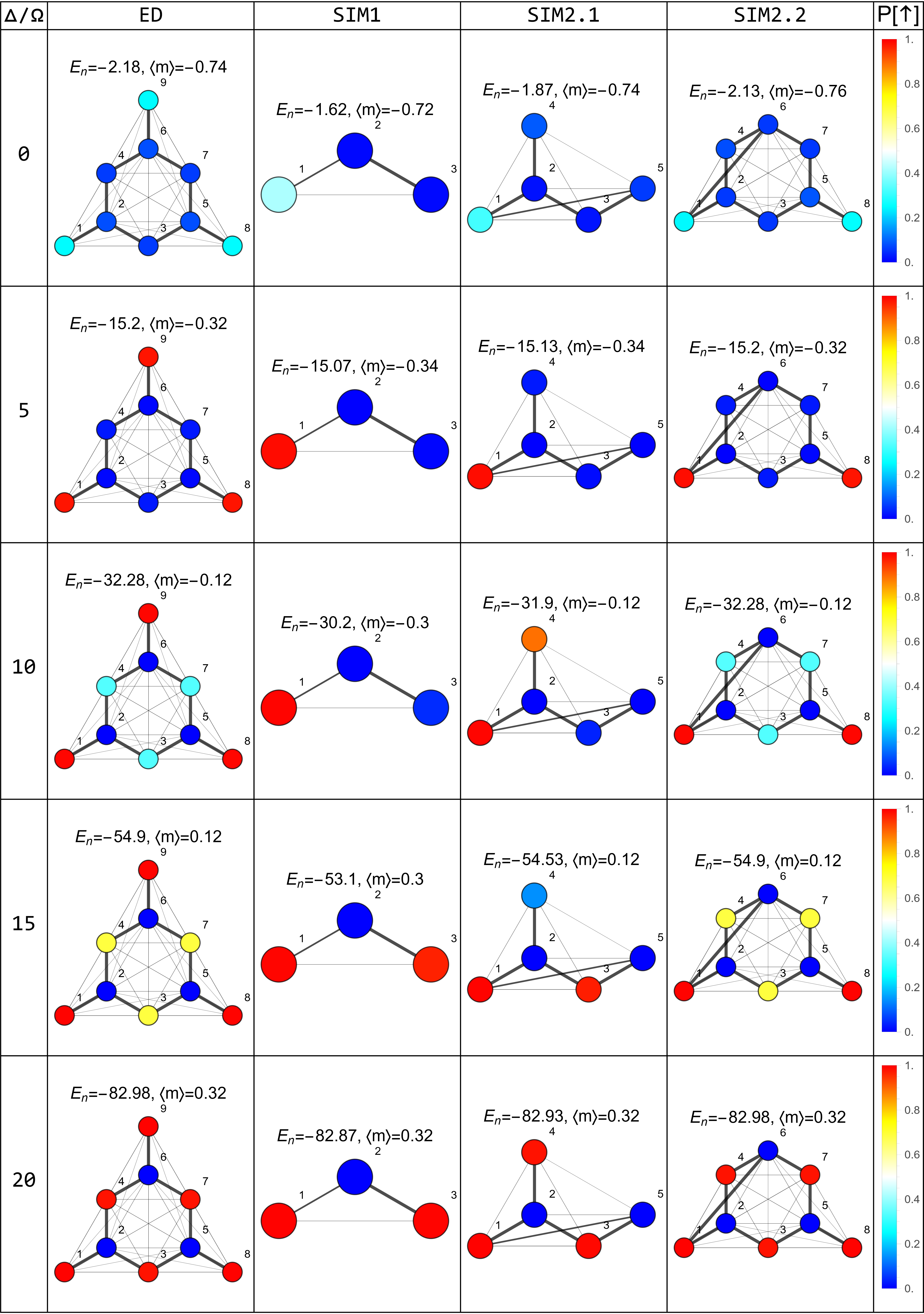}
    \caption{The local ground state magnetization resulting from the various \sg methods is presented and compared for the first generation Sierpinski triangle with VdW interactions at $V/\Omega=110$, the same value studied in Ref.~\cite{verstraten2025control}. Remark that SIM2 has an important feature that is not visible but is necessary for this comparison, namely that added sites have a different ``weight" compared to symmetric sites. In particular here,  site 3 has twice the weight compared to site 4. That is, site 4 is a singular site added from the NN, while site 3 also represents site 7 in the original geometry. }
    \label{fig:sim2states}
\end{figure}

Let us finally consider the local magnetizations, which feature in the geometry of the \sg methods. In Fig.~\ref{fig:sim2states}, we compare the local groundstate magnetizations of the first generation Sierpinski triangle for the same interaction strength as in Fig.~\ref{fig:sierpsim2}, spaced every five $\Delta/\Omega$. We saw that SIM1 missed the two intermediate plateaus, which is visible in Fig.~\ref{fig:sim2states} when we compare site 3 between ED and SIM1 for the cases $\Delta/\Omega=10$ and $15$. Due to the 3 fold symmetry, SIM1 jumps the number of edge excitations straight from 0 to 3 right at the transition where the number of edge excitations should go from 1 to 2, according to ED. Considering SIM2.1, we first need to remark a difference between sites 3 and 4, which is not visually indicated.There is actually a single weight on site 4, while there is a double weight on site 3, since SIM2.1 has taken one weight away from the symmetric subset and moved it to a NN. The same holds for sites 2 and 5. Now that we know this technical difference, we can see that all working cases from SIM1 remain unchanged. However, when we consider $\Delta/\Omega=10$ and $15$, we see some wrong colors. This is to be expected, however, due to these differing weights. When there is only 1 spin in a superposition, SIM2.1 puts a spin on the site with weight 1. Then, when there are two spins on the edges, the spin moves to site 3, doubling in weight since this represents site 7 as well (remember that, in SIM2.1, all sites that are not NN are projected to the subset defined by SIM1).  In other words, by splitting up the sites into two subclasses, we manually break the symmetry, but we gain the ability to describe the various superpositions by their subclass. Finally, in SIM2.2 we nearly recover the original geometry, apart from a single corner which is projected to site 1. Although this might appear trivial, the reduction of a single site is nevertheless a factor $2^2\sim 2^3$ increase of computational speed, depending on the method.

\section{Conclusions}
\label{sec:con}

In this paper, we have demonstrated how to apply the new SIM-GRAPH method to any finite-size quantum Ising system. We have compared the graph method to the direct matrix manipulations, and showed that the symmetry reduced graph is effectively the Hilbert subspace of symmetric configurations, which tends~\footnote{We do note that topological properties may not hold to these standard expectations. Indeed, long-range entanglement is one of the first properties projected away in SIM-GRAPH.} to be the most likely class for the ground-state. Since this is a projective method, the computed eigenenergy will always be equal to or greater than the true eigenenergy. Thus, it is not guaranteed that we always find the ground state, but we will always find the lowest energy symmetric state, which can never have a magnetization that is different from the ground state by more than the symmetry factor. This makes SIM-GRAPH a very reliable method for computing the ground-state magnetization of large-size systems. Furthermore, we have introduced two extra modifications to the standard method, which allow for even geometries and partial symmetry-broken states to be included into the formalism. That is, SIM2 includes neighbors of the symmetric subset into the reduction, allowing the most likely superpositions to be included in the subspace. On the other hand, SIM3 only uses a subgroup of the full symmetry group, which allows for symmetry-broken states in the rest of the subclass. In other words, one may study reflection broken states, while still using the rotational symmetry for the reduction.

The number of sites $N$ is typically the most important factor in computational expense, both in terms of memory ($\sim 2^N$ bytes required per state) and computational time ($\sim 2^{3N}$ for full ED). Since SIM-GRAPH tends to reduce $N$ by a factor on the order of the symmetry class $S$, we gain significant exponential improvements compared to ED. Furthermore, the resulting effective Hamiltonian may still be solved using other numerical techniques. Thus, there is no reason not to combine other methods with SIM-GRAPH when you only require groundstate observables. It will either be exponentially faster, or one can compute properties of systems on the order of $S$ times the number of sites. For a square geometry, that is a potential increase of the size limit by a factor of $\sim8$.

An interesting application of \sg is to combine the strengths of different methods. Although \sg may slightly misjudge the value of a plateau, it is often very accurate at finding the transition points between plateaus. Thus, combining a high-resolution computation from \sg with low-resolution exact results may yield surprisingly high accuracy without requiring as much computational time as using a single method. We also believe that SIM-GRAPH may be used in bio-chemistry, where one often deals with very large molecules that obey certain symmetries~\cite{li2023structural}. While these may not be described by Ising models, the basic principles of SIM-GRAPH only require the ability to describe the system as local Hilbert spaces, which are all interconnected via interactions. This allows for the graph description, analysis, projection, and reconstruction of an effective system. Thus, the ``I" in \sg could be replaced by Interacting instead of Ising.

\section{Acknowledgments}
We thank Fabien Alet for interesting discussions.
This work was supported by the Netherlands Organization
for Scientific Research (NWO, Grant No. 680.92.18.05,
C.M.S. and R.C.V.)

% \appendix
% \section{Review on Fractional Calculus}
% \label{AppFC}
%\clearpage

\bibliography{apssamp}% Produces the bibliography via BibTeX.

\end{document}